\documentclass{iopart}

\usepackage{iopams}
\usepackage{graphicx}
\usepackage{color}

\newcommand{\bra}[1]{\ensuremath{\langle #1 |}}
\newcommand{\ket}[1]{\ensuremath{| #1\rangle}}
\newcommand{\bk}[2]{\ensuremath{\langle #1 | #2\rangle}}
\newcommand{\kb}[2]{\ensuremath{| #1\rangle\!\langle #2 |}}
\newcommand{\Sr}{\mathrm{Sr}}
\newcommand{\Conv}{\mathrm{Conv}}

\newcommand{\ot}{\otimes}
\newcommand{\ti}{\times}

\newcommand{\cD}{{\cal D}}
\newcommand{\cA}{{\cal A}}
\newcommand{\cL}{{\cal L}}
\newcommand{\cJ}{{\cal J}}
\newcommand{\cS}{{\cal S}}

\newcommand{\cB}{{\cal B}}
\newcommand{\cC}{{\cal C}}

\newcommand{\cO}{{\cal O}}
\newcommand{\cP}{{\cal P}}

\newcommand{\cE}{{\cal E}}

\newcommand{\cX}{{\cal X}}
\newcommand{\cH}{{\cal H}}
\newcommand{\cW}{{\cal W}}
\newcommand{\cV}{{\cal V}}
\newcommand{\la}{\langle}
\newcommand{\ra}{\rangle}

\newcommand{\R}{{\mathbf R}}
\newcommand{\C}{{\mathbb C}}

\newcommand{\wt}{\widetilde}

\newcommand{\bI}{\mathbb{I}}
\newcommand{\wJ}{\wt{\cJ}}

\mathchardef\zb="710C  
\mathchardef\zg="710D  
\mathchardef\zd="710E  
\mathchardef\zve="710F 
\mathchardef\zz="7110  
\mathchardef\zh="7111  
\mathchardef\zvy="7112 
\mathchardef\zi="7113  
\mathchardef\zk="7114  
\mathchardef\zl="7115  
\mathchardef\zm="7116  
\mathchardef\zn="7117  
\mathchardef\zx="7118  
\mathchardef\zp="7119  
\mathchardef\zr="711A  
\mathchardef\zs="711B  
\mathchardef\zt="711C  
\mathchardef\zu="711D  
\mathchardef\zvf="711E 
\mathchardef\zq="711F  
\mathchardef\zc="7120  
\mathchardef\zw="7121  
\mathchardef\ze="7122  
\mathchardef\zy="7123  
\mathchardef\zf="7124  
\mathchardef\zvr="7125 
\mathchardef\zvs="7126 
\mathchardef\zf="7127  
\mathchardef\zG="7000  
\mathchardef\zD="7001  
\mathchardef\zY="7002  
\mathchardef\zL="7003  
\mathchardef\zX="7004  
\mathchardef\zP="7005  
\mathchardef\zS="7006  
\mathchardef\zU="7007  
\mathchardef\zF="7008  
\mathchardef\zW="700A  
\newcommand{\ee}{\end{equation}}

\newcommand{\be}{\begin{equation}}
\newcommand{\bea}{\begin{eqnarray}}
\newcommand{\eea}{\end{eqnarray}}
\newcommand{\beas}{\begin{eqnarray*}}
\newcommand{\eeas}{\end{eqnarray*}}
\newcommand{\nm}[1]{\ensuremath{\Vert #1 \Vert}}

\newtheorem{theo}{Theorem}[section]
\newtheorem{prop}{Proposition}[section]
\newtheorem{lem}{Lemma}[section]
\newtheorem{cor}{Corollary}[section]

\newtheorem{problem}{Problem}[section]

\begin{document}

\title{Convex bodies of states and maps}

\author{Janusz Grabowski$^1$, Alberto Ibort$^2$\footnote{On leave of absence from
Depto. de Matem\'aticas, Univ. Carlos III de Madrid, 28911 Legan\'es, Madrid,
Spain.}, Marek Ku\'s$^3$ and Giuseppe Marmo$^4$}

\address{$^1$Polish Academy of Sciences, Institute of Mathematics,\'Sniadeckich 8, P.O. Box 21, 00-956 Warsaw,
Poland}

\address{$^2$Departament of Mathematics, University of California at Berkeley, Berkeley, CA 94720,
USA}

\address{$^3$Center for Theoretical Physics, Polish Academy of Sciences, Aleja Lotnik{\'o}w 32/46, 02-668 Warszawa,
Poland}

\address{$^4$Dipartimento di Fisica, Universit\`{a} ``Federico II'' di Napoli and Istituto Nazionale di Fisica Nucleare,
Sezione di Napoli, Complesso Universitario di Monte Sant Angelo, Via Cintia,
I-80126 Napoli, Italy}

\eads{\mailto{jagrab@impan.pl}, \mailto{albertoi@math.uc3m.es},
\mailto{marek.kus@cft.edu.pl}, \mailto{marmo@na.infn.it}}



\begin{abstract} We give a general solution to the question when the
convex hulls of orbits of quantum states on a finite-dimensional Hilbert space
under unitary actions of a compact group have a non-empty interior in the
surrounding space of all density states. The same approach can be applied to
study convex combinations of quantum channels. The importance of both problems
stems from the fact that, usually, only sets with non-vanishing volumes in the
embedding spaces of all states or channels are of practical importance. For the
group of local transformations on a bipartite system we characterize maximally
entangled states by properties of a convex hull of orbits through them. We also
compare two partial characteristics of convex bodies in terms of largest balls
and maximum volume ellipsoids contained in them and show that, in general, they
do not coincide. Separable states, mixed-unitary channels and $k$-entangled
states are also considered as examples of our techniques.


\end{abstract}

\section{Introduction}
In many issues of quantum information theory and geometry of quantum states,
one is confronted with the problem whether some subset of states or quantum
channels is `large enough' to be of significance in applications. On a
qualitative level the problem can be reduced to the question whether the
considered set contains an open subset (as a subset of the set of all
states/channels). If the answer is affirmative, one can ask more quantitative
questions about the relative volume of such sets, about some estimates of their
volumes, or radiuses of the maximal balls they contain.Many questions of this
type can be regarded as instances of the following general problem (see
e.g.~\cite{barvinok05}).

\begin{problem}\label{pr1}
Let $\mathcal{V}$ be an Euclidean space, i.e.\ a finite-dimensional real
vector space equipped with a scalar product $\bk{\cdot}{\cdot}_\cV$ and let
$K$ be a compact group acting on $\mathcal{V}$ by orthogonal transformations,
$K\times\cV\ni(U,x)\mapsto U \cdot x\in\cV$. Given a $K$-invariant affine subspace
$\cA$ of $\cV$ and a vector $x_0\in\cA$, decide whether the convex hull
$\Conv(K\cdot x_0)$ of the orbit $K \cdot x_0$ (the \emph{convexed orbit}) is a {\it
convex body} in $\cA$, i.e.\ whether the interior of $\Conv(K\cdot x_0)$ is a
non-empty open set of $\cA$. This is the same as to decide whether the volume
of $\Conv(K\cdot x_0)$ is positive in $\cA$.
\end{problem}


Let us recall that affine subspaces in $\cV$ are exactly subsets closed with
respect to \emph{affine combinations},

\[
a_1,a_2\in\cA\ \Rightarrow\ \forall\, t\in\R\ [ta_1+(1-t)a_2\in\cA]\,,
\]
and that the differences $a_1-a_2$ of points of $\cA$ form a real vector
subspace $V(\cA)$ of $\cV$ called the \emph{linear part of $\cA$}. {\it Convex
combinations} are those affine combinations $ta_1+(1-t)a_2$ for which $0\le
t\le 1$. After recalling in \Sref{sec:notations} some elementary notions, we
present in \Sref{sec:examples} several examples to which our analysis can be
applied, both in the cases of states and channels. Some of them concern
problems for which the answer to the posed question is known, but they provide
a perfect insight into a unifying power of our approach.The full answer to the
above stated Problem~\ref{pr1} is given in \Sref{sec:solution}. In
\Sref{sec:applications} we show how to apply the obtained result to the
examples of \Sref{sec:examples}. In the case of maximally entangled states our
approach leads to a unique characterization of such states in terms of
properties of convexed orbits through them. In \Sref{sec:ellipsoids} we compare
characterizations of convex bodies of states in terms of the largest ball which
can be inscribed within the body in question and, so called, the \emph{maximal
volume ellipsoid} of that body. In principle, the later notion is an affine
one, whereas the former bears a metric nature. However, in some important cases
both notions coincide (e.g.\ for the set all density states), in other (e.g.\
for convexed local orbits of pure states in composite systems) this is no
longer true.

\section{Notations and conventions}\label{sec:notations}
Let $\cH$ be an $n$-dimensional Hilbert space with a Hermitian product $\la
x,y\ra_\cH$ being, by convention, $\C$-linear with respect to $y$ and
anti-linear with respect to $x$. Let $gl(\cH)$ be the complex vector space of
all complex linear operators on $\cH$. It is also canonically a Hilbert space
with the Hermitian product \be \label{metric0} \la A,B\ra_{gl}=\tr(A^\dag B)\,,
\ee where $A^\dag$ is the Hermitian conjugate of $A$, i.e., $ \la
Ax,y\ra_\cH=\la x,A^\dag y\ra_\cH$.The unitary group $U(\cH)$ consists of those
complex linear operators $U\in gl(\cH)$ on $\cH$ which satisfy $UU^\dag=I$. It
acts canonically on $\cH$ preserving the Hermitian product.Fixing an
orthonormal basis $(e_k)$ of $\cH$ allows us to identify the Hermitian product
$\la x,y\ra_\cH$ on $\cH$ with the canonical Hermitian product on $\C^n$ of the
form $\la a,b\ra_{\C^n}=\sum_{k=1}^n\overline{a_k}{b_k}$, the group $U(\cH)$ of
unitary transformations of $\cH$ with $U(n)$, its Lie algebra $u(\cH)$ with
$u(n)$, etc. In this picture, $(a_{jk})^\dag=(\overline{a_{kj}})$.One important
convention we want to introduce is that we identify the (real) vector space of
Hermitian operators with the dual $u^*(\cH)$ of the (real) Lie algebra
$u(\cH)$, according to the pairing between Hermitian, $A\in u^*(\cH)$, and
anti-Hermitian, $T\in u(\cH)$, operators: $\la A,T\ra=\tr(AT)$. The
multiplication by $i$ establishes further a vector space isomorphism $u(\cH)\ni
T\mapsto \rmi T\in u^*(\cH)$ which identifies the adjoint and the coadjoint
action of the group $U(\cH)$, $\mathrm{Ad}_U(T)=U T U^\dag$. Under this
isomorphism, $u^*(\cH)$ becomes a Lie algebra with the Lie bracket
$[A,B]=\frac{1}{\rmi}(AB-BA)$, equipped additionally with the scalar product

\be\label{metric1}\la A,B\ra_{u^*}=\tr(AB) \ee
and the corresponding
Hilbert-Schmidt (Frobenius) norm $\nm{A}_{u^*}=\sqrt{\tr(A^2)}$.

\section{Examples}\label{sec:examples}

\subsection{Density states}

The space of all non-negatively defined operators, i.e.\ of those $\zr\in
gl(\cH)$ which can be written in the form $\zr=T^\dag T$ for a certain $T\in
gl(\cH)$, we denote by $\cP(\cH)$. It is a convex cone in the Euclidean space
$\cV=u^*(\cH)$. The set of {\it density states}, $\cD(\cH)$, is distinguished
in the cone $\cP(\cH)$ by the equation $\tr(\zr)=1$, so it is a convex subset
in the affine subspace $\cA=u^*_1(\cH)\subset u^*(\cH)$ of trace 1 Hermitian
operators. The linear part of $u^*_1(\cH)$ is the subspace
$u^*_0(\cH)=su^*(\cH)$ of Hermitian operators with trace 0.  Denote by
$\cD^k(\cH)$ the set of all density states of rank $k$. In the standard
terminology, $\cD^1(\cH)$ is the space of \emph{pure states}, i.e.\ the set of
one-dimensional orthogonal projectors $\mid \psi\ra\la \psi\mid$, where $\Vert
\psi\Vert^2=\bk{\psi}{\psi}=1$. It is known that the set of extreme points of
$\cD(\cH)$ coincides with the set $\cD^1(\cH)$ of pure states. Hence, every
element of $\cD(\cH)$ is a convex combination of points from $\cD^1(\cH)$. The
space $\cD^1(\cH)$ of pure states can be identified with the complex projective
space $\mathbb{P}(\mathcal{H})\simeq\C P^{n-1}$ via the projection

\[
\cH\setminus\{ 0\}\ni \psi\mapsto P_\psi=\frac{\mid\! \psi\ra\la
\psi\!\!\mid}{\Vert \psi\Vert^2}\in\cD^1(\cH)
\]
which identifies the points of the orbits of the $\C\setminus\{ 0\}$-group
action by complex homoteties. Actually, due to the probabilistic
interpretation, a pure quantum state is a point in this projective space
$\mathbb{P}(\mathcal{H})\simeq \cD^1(\cH)$ rather than a vector in
$\mathcal{H}$.The unitary group $K=U(\cH)$ acts canonically and orthogonally on the
Euclidean space $\cV=u^*(\cH)$ by \be\label{action}A\mapsto U.A=UA U^\dag=UA
U^{-1}\,,\ee and the orbits of this action are distinguished by the spectrum
of  the Hermitian operator $A$. Of course, we can consider the
$U(\cH)$-action on the Hilbert space $gl(\cH)$ as the complexification of the
orthogonal action on $u^*(\cH)$, since

\[
gl(\cH)=\C\otimes u^*(\cH)=u^*(\cH)\oplus \rmi u^*(\cH)=u^*(\cH)\oplus
u(\cH)\,.
\]
All operators proportional to the identity, $\zl I$, are fixed points of this
action. It is also easy to see that the trace is preserved, so that the affine
spaces $u^*_\lambda(\cH)=\{ A\in u^*(\cH):\tr A=\lambda\}$ are invariant under
the $U(\cH)$-action. In particular, for any $\ket{\psi}\in\cH$, $\ket{\psi}\ne
0$, the orbit $U(\cH).P_{\psi}$ is a minimal orbit of $U(\cH)$ in
$\cA=u^*_1(\cH)$ which coincides with the set $\cD^1(\cH)$ of pure states and
whose convex hull $\Conv(U(\cH).P_{\psi})$ is the convex set $\cD(\cH)$ of all
(mixed) states.It is well known that $\cD^1(\cH)$ is canonically a K\"ahler
manifold with respect to the metric induced from $u^*(\cH)$, the {\it
Fubini-Study metric}, and the symplectic form of a coadjoint orbit of $U(\cH)$
(cf.~\cite{grabowski05}).

\subsection{States of composite systems}

The Hilbert space of a bipartite composite system is the tensor product of
subsystem Hilbert spaces,

\begin{equation}\label{Hcomp}
\mathcal{H}=\mathcal{H}_1\otimes\mathcal{H}_2.
\end{equation}
A pure state in $\mathcal{H}$ is \emph{separable} if it corresponds to a simple
tensor,
\begin{equation}\label{pureseparable}
\ket{\psi}=\ket{\phi^1}\otimes\ket{\phi^2}.
\end{equation}
As such, it can be identified with the rank-one projection,
\begin{equation}\label{puredensity}
P_\psi=\frac{\kb{\psi}{\psi}}{\bk{\psi}{\psi}}.
\end{equation}

Denote the set of separable pure states with
$\cS^1(\cH)=\cS^1(\cH_1\otimes\cH_2)$ (this depends on the decomposition of
$\cH$ into the tensor product). It is easy to see that it is a single minimal
orbit $\cO_{P_{\phi^1}\otimes P_{\phi^2}}$ of the obvious orthogonal action
of $K=U(\cH_1)\times U(\cH_2)\subset U(\cH_1\otimes\cH_2)$ on the Euclidean
space

\[
\cV=u^*(\cH_1\otimes\cH_2)=u^*(\cH_1)\ot u^*(\cH_2)
\]
going through the point
$P_{\phi^1}\otimes P_{\phi^2}$ for some (arbitrary)
$\ket{\phi^{1,2}}\in\mathcal{H}_{1,2}$,

\begin{equation}\label{separableorbit}
\cS^1(\cH)=\{(U_1P_{\phi^1}U_1^\dagger)\otimes (U_2P_{\phi^2}U_2^\dagger):\, U_i\in U(\mathcal{H}_i), \, i=1,2\}.
\end{equation}
A mixed state $\rho$ is, by definition, \emph{separable} if it belongs to the
convex hull of this orbit, i.e.\ it is a convex combination of pure separable
states,

\begin{equation}\label{mixedseparable}
\rho=\sum_{k=1}^n p_k P_{\phi^1_k}\otimes P_{\phi^2_k},
\quad p_k\ge 0,\quad \sum_{k=1}^n p_k=1,
\end{equation}
for some $\phi^i_1,\ldots,\phi^i_n\in\mathcal{H}_i$, $i=1,2$. The other states
are called {\it entangled}.The problem whether the set $\cS(\cH)$ of mixed
separable states possesses a nonzero volume, (cf.~\cite{zyczkowski98}), reduces
to the question whether $\Conv(\cS^1(\cH))$ contains a non-trivial open subset
of $\cA=u^*_1(\cH_1\otimes\cH_2)\subset\cV$.It is known that any element
$\ket{\psi}\in\cH_1\ot\cH_2$ admits a \emph{Schmidt decomposition}

\be\label{Sd}\ket{\psi}=\sum_{j=1}^r\lambda_j\cdot\ket{\phi^1_j}\otimes\ket{\phi^2_j}\,,
\ee with $(\ket{\phi^1_j})$ and $(\ket{\phi^2_j})$ being (not necessarily
complete) orthonormal sets, and $\zl_j$ being positive real numbers. The number
$r$ of summands in this decomposition we call the {\it Schmidt rank} of
$\ket{\psi}$ and denote $\Sr(\psi)$. Directly by definition, a pure state
$P_\psi=\ket{\psi}\bra{\psi}$ on $\cH_1\ot\cH_2$ is {\it separable} if and only
if the Schmidt rank of $\ket{\psi}$ is 1.This easy characterization of
separable pure states has been used by Terhal and Horodecki \cite{terhal00} to
develop the concept of {\it Schmidt number} of an arbitrary density state $\zr$
(quantum state in finite dimensions). This number characterizes the minimum
Schmidt rank of the pure states that are needed to construct such density
matrix. The Schmidt number is non-increasing under local operations and
classical communications, i.e.\ it provides a legitimate entanglement measure.
We can construct an entanglement measure, the {\it Schmidt measure $\zm_S$},
which is additionally convex, using the convex roof construction (see e.g.\
\cite{eisert01}). This construction, proposed as a general tool for
entanglement measures (see e.g.\ \cite{grabowski05,uhlmann00,grabowski06}), can
be repeated in infinite dimensions as

\begin{equation}\label{S}
\zm_S(\zr)=\inf \left\{\sum_jp_j \Sr(\zc_j)\right\}\,,
\end{equation}
where the {\it infimum} is taken over all possible realizations of $\zr$ as
infinite-convex combinations $\zr=\sum_jp_j\ket{\zc_j}\bra{\zc_j}$ with $0\le
p_j\le 1$, $\sum_jp_j=1$ and $\ket{\zc_j}\in\cH_1\ot\cH_2$. Every quantum state
admits such a realization and a reasoning analogous to the one in
\cite{grabowski05} shows that $\zm_S$ is infinite-convex, non-negative, and
vanishes exactly on separable states.The Schmidt rank can be conveniently
expressed in terms of the \emph{Jamio{\l}kowski isomorphism}

\[
\wt{\cJ}:\cL\left(gl(\cH_2), gl(\cH_1)\right)\to gl(\cH_1\ot\cH_2)\,,
\]
identifying linear maps on $\cH_1\ot\cH_2$ with the space
$\cL\left(gl(\cH_2), gl(\cH_1)\right)$ of linear maps $\Phi:gl(\cH_2)\to
gl(\cH_1)$ as follows.

\begin{theo}\cite{grabowski07}
The Schmidt rank of $\ket{\psi}$ is $r$ if and only
if $\wt{\cJ}^{-1}(P_\psi):gl(\cH_2)\to gl(\cH_1))$ is a linear operator
of rank $r^2$. In particular, $P_\psi$ is separable if and only if $\wt{\cJ}^{-1}(P_\psi)$ is of rank 1.
\end{theo}

Recall that a pure state $P_\psi$ we call \emph{$k$-entangled} if the Schmidt
rank of $\ket{\psi}$ is $\le k$. Denote the family of all such states with
$\cE_k(\cH_1\ot\cH_2)$. This concept emerged from the study of a duality for
{\it $k$-positive maps} \cite{terhal00,skowronek09,szarek10}. According to
the above theorem, $P_\psi\in\cE_k(\cH_1\ot\cH_2)$ if and only if
$\wt{\cJ}^{-1}(P_\psi):gl(\cH_2)\to gl(\cH_1))$ is a linear operator of rank
$\le k^2$. A mixed state $\zr$ on $\cH_1\ot\cH_2$ is called
\emph{$k$-entangled} if it belongs to the convex hull
$\Conv(\cE_k(\cH_1\ot\cH_2))$. Note that 1-entangled states are exactly
separable states.

\subsection{Maximally entangled states}

If we assume that $\dim(\cH_1)\ge\dim(\cH_2)=m$, then the Schmidt rank of any
$\ket{\psi} \in\cH_1\ot\cH_2$ is not bigger than $m$. Moreover, $\Sr(\psi)=m$
if and only if

\[
\ket{\psi}=\sum_{j=1}^{m}\lambda_j\cdot\ket{\phi^1_j}\otimes\ket{\phi^2_j}\,,
\]
where $(\ket{\phi^2_j})$ is an orthonormal basis in $\cH_2$,
$(\ket{\phi^2_j})$ is an orthonormal system in $\cH_1$, and $\lambda_j>0$,
$j=1,\dots,m$. The corresponding pure state $P_\psi$ is called
\emph{maximally entangled} if all $\lambda_j$ are equal, i.e.\ for normalized
$\ket{\psi}$, $\lambda_j=1/\sqrt{m}$, $j=1,\dots,m$. Since, for normalized
$\ket{\psi}$, $1=\bk{\psi}{\psi}=\sum_j\lambda_j^2$,

\[
P_\psi=\kb{\psi}{\psi}=\sum_{i,j=1}^{m}\lambda_i\lambda_j\kb{\phi^1_j}{\phi^1_i}\otimes\kb{\phi^2_j}{\phi^2_i}\,,
\]
the (obviously defined) \emph{partial traces} are
\begin{equation}\label{partialtrace}
\tr_1P_\psi=\sum_{j=1}^{m}\lambda^2_j\kb{\phi^2_j}{\phi^2_j}\,,\quad \tr_2P_\psi=\sum_{j=1}^{m}\lambda^2_j\kb{\phi^1_j}{\phi^1_j}\,.
\end{equation}

It follows that $P_\psi$ is maximally entangled if and only if $\tr_1P_\psi$
is proportional to the identity operator ${I}_{\cH_2}$ on $\cH_2$,

\be\label{maxent}\tr_1P_\psi=\mathbb{I}_{\cH_2}\,, \ee where
$\mathbb{I}_{\cH}=\frac{1}{\dim(\cH)}I_\cH$. Moreover,
$\tr_2P_\psi=\mathbb{I}_{\cH_1}$ if and only if $\dim(\cH_1)=\dim(\cH_2)$ and
$\tr_1P_\psi=\mathbb{I}_{\cH_2}$. Conversely, if (\ref{maxent}) is satisfied,
then, in view of (\ref{partialtrace}),

\[
\sum_{j=1}^{m}\lambda^2_j\kb{\phi^2_j}{\phi^2_j}=\frac{1}{m}I_{\cH_2}\,,
\]
so $\lambda_1=\cdots=\lambda_{m}=\frac{1}{\sqrt{m}}$ and we get the following.

\begin{prop}\label{p0} A pure state $P_\psi$ on $\cH_1\ot\cH_2$, $\dim(\cH_1)\ge\dim(\cH_2)>1$, is maximally entangled if and only if $\tr_1P_\psi=\bI_{\cH_2}$. Moreover, $\tr_2P_\psi=\mathbb{I}_{\cH_1}$ if and only if $\dim(\cH_1)=\dim(\cH_2)$ and $P_\psi$ is maximally entangled.
\end{prop}

From the above it is clear that the $K=U(\cH_1)\times U(\cH_2)$-orbit $\mathcal{O}_{\zr_{max}}$ through a maximally entangled state $\zr_{max}$,

\begin{equation}\label{maxentorbit}
\mathcal{O}_{\zr_{max}}
=\{(U_1\otimes U_2)\circ\rho_{max}\circ (U_1^\dagger\otimes U_2^\dagger):\,
 U_i\in U(\mathcal{H}_i), \, i=1,2\}\,,
\end{equation}
consists of all maximally entangled pure states. We can ask whether the convex
hull of this orbit is a convex body in the affine space
$\cA=u^*_1(\cH_1\otimes\cH_2)$. Although the problem \textit{per se} might be
not of a particular interest, it is closely related (by the Jamio{\l}kowski
isomorphism) to that of Example \ref{uc} below which draws much attention.

\subsection{Mixed-unitary channels}\label{uc}

Let, as before, $\mathcal{H}$ be a finite-dimensional Hilbert space. In the
simplest setting, a {\it quantum channel} or a {\it stochastic map} is a
completely positive, trace preserving map $A:gl(\mathcal{H})\mapsto
gl(\mathcal{H})$. According to the Choi's theorem, any completely positive
map can be written in the form of a {\it Kraus map}
\begin{equation}\label{kraus}
A(\rho)=\sum_kX_k\rho X_k^\dagger, 
\end{equation}
for some $X_k\in gl(\mathcal{H})$. To ensure trace preserving, they have to
fulfil
\be\label{stoch}\sum_kX_k^\dagger X_k={I_\cH}\,. \ee
One considers also {\it doubly stochastic channels} for which not only the trace but also
the identity is preserved, $A(I_\cH)=I_\cH$, i.e.,

\be\label{ds} \sum_kX_k
X_k^\dagger={I_\cH}\,.
\ee

Let us point out that the $\mathbb{R}$-linear span of Kraus maps is the space
$HP(gl(\cH))$ of {\it Hermiticity preserving} operators $A:gl(\cH)\to gl(\cH)$.
On this space there are two natural maps $T_1,T_2:HP(gl(\cH))\to u^*(\cH)$
defined on Kraus maps (\ref{kraus}) by

\be\label{T}T_1(A)=\sum_kX_k^\dagger X_k\,,\quad T_2(A)=\sum_kX_k
X_k^\dagger\,, \ee
so that a doubly stochastic channel is a completely
positive map $A$ satisfying $T_1(A)=T_2(A)=I_\cH$.
The set we want to investigate from our general point of view is the set
$\cC_{MUC}$ of \textit{mixed-unitary channels} \cite{watrous08}, consisting
of doubly stochastic channels of the form

\begin{equation}\label{mixunitchannel}
A(\rho)=\sum_kp_kU_k\rho U_k^\dagger, \quad U_kU_k^\dagger={I_\cH}, \quad p_k>0,
\quad \sum_kp_k=1.
\end{equation}

This is clearly the convex hull of the set of doubly stochastic channels
$\{\zr\to U\zr U^\dag: U\in U(\cH\}$ which can be interpreted also as the
orbit $\cO_{MUC}$ of the identity channel $I_{gl}(\zr)=\zr $ under the group
$K=U(\cH)\ti U(\cH)$ acting on the Hilbert space $gl(gl(\cH))$ by

\be\label{uaction}((U_1,U_2).A)(\zr)=U_1A(U_2\zr U_2^\dag) U_1^\dag\,. \ee

Under the identification {\it via} the Jamio{\l}kowski isomorphism
\cite{jamiolkowski72,choi75,grabowski07}

\be\label{jam}\wt{\cJ}:gl(gl(\cH))\to gl(\cH\ot\cH)\,, \ee the real vector
space $HP(gl(\cH))$ of Hermiticity preserving maps corresponds to the
Euclidean space $\cV=u^*(\cH\ot\cH)=u^*(\cH)\ot u^*(\cH)$ of Hermitian
operators on $\cH\ot\cH$, and completely positive maps to non-negatively
defined operators. With this identification, the $K=U(\cH)\ti U(\cH)$-action
(\ref{uaction}) goes to the obvious tensor product $K$-action,

\be\label{newaction}(U_1,U_2).(X_1\ot
X_2)=(U_1X_1U_1^\dag)\ot(U_2X_2U_2^\dag)\,. \ee
The question how big is $\cC_{MUC}$ is therefore equivalent to the question how big is $\Conv(\cO)$ for the orbit $\cO=K.\wJ(I_{gl})$. We will come back to this problem in section \ref{muc}.

\section{Characterizing convex bodies}\label{sec:solution}

\subsection{A solution}
An answer to Problem~\ref{pr1}  is given by the following.

\begin{theo}\label{th:main} Under assumptions of Problem~\ref{pr1}, the convex hull
$\Conv(K\cdot x_0)$ has empty interior in $\cA$ if and only if there exists a proper
invariant subspace ${\cW}$ of the linear part $\cX={V}(\cA)$ such that
$x_0\in{\cV}_K+{\cW}$, where ${\cV}_K=\{x\in{\cV}:\,
K.x=x\}$ is the subspace of $K$-stationary points.
\end{theo}
\textbf{Proof} Let us assume that $\Conv(K\cdot x_0)$ has empty interior in
$\cA$. It means that it is contained in a proper affine subspace
$\mathcal{A}_0$ of $\mathcal{A}$; $\cA_0$ is the affine span of $\Conv(K\cdot
x_0)$, $\mathcal{A}_0=\mathit{Aff}(K.x_0)$. The affine subspace $\mathcal{A}_0$
is invariant with respect to the action of $K$,
$K.\mathcal{A}_0\subset\mathcal{A}_0$, and the same is true for its linear part
$V(\mathcal{A}_0)=\cX$. Since the action of $K$ is orthogonal, the orthogonal
complement $\mathcal{X}^\perp$ is invariant as well. Due to a dimensional
argument, $\mathcal{X}^\perp$ and $\mathcal{A}$ intersect at a single point
$v\in\mathcal{X}^\perp\cap\mathcal{A}$. Since both $\mathcal{X}^\perp$ and
$\mathcal{A}$ are $K$-invariant, it follows that $K.v=v$, i.e.,
$v\in\mathcal{V}_K$. But then $x_0=v+x^\prime$ for some
$x^\prime\in\mathcal{X}$, so we can take $\mathcal{W}=\mathcal{X}$.Let us now
assume that $x_0=v+w$ for some $v\in\mathcal{V}_K$ and $w\in\mathcal{W}$, where
$\mathcal{W}$ is a proper invariant subspace of $\cV$. Then, the orbit
$K.x_0=v+K.w$ is contained in the proper affine subspace $v+\mathcal{W}$ of
$\cA$, hence it has empty interior in $\cA$. \hfill $\square$

\begin{figure}[h]
\begin{center}
\includegraphics[scale=.5]{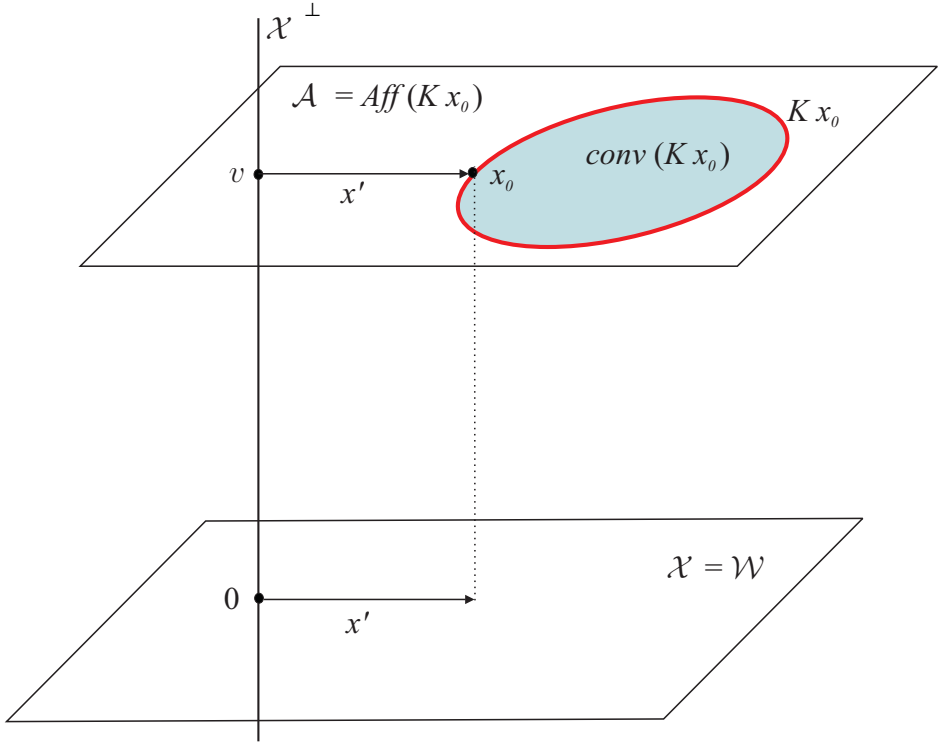}
\caption{Illustration of Theorem~\ref{th:main} (see text for notation and explanation)}
\end{center}
\end{figure}

\begin{cor}\label{cmain}
The convex hull $\Conv(K\cdot x_0)$ is a convex body in $\cA$ if and only if
the image of $x_0$ under the orthogonal projection $\zp:\cV\to V(\cA)$ does not
belong to a proper $K$-invariant subspace of $V(\cA)$. In particular, if
$V(\cA)$ is irreducible and $\pi(x_0)\ne 0$, then $\Conv(K\cdot x_0)$ is a
convex body in $\cA$.
\end{cor}
\textbf{Proof} There is a unique vector $v\in\cV(\cA)^\perp$ such that
$\cA=v+\cV(\cA)$, thus $v\in\cV_K$ due to the invariance of $\cA$ and
$\cV(\cA)$. Hence, $\zp(x_0)$ belongs to a proper invariant
$\cW\subset\cV(\cA)$ if and only if $x_0\in v+\cW$, so $\Conv(K\cdot x_0)$ has
empty interior in $\cA$ due to the above theorem. \hfill $\square$

\subsection{Convexity and coadjoint orbits}

A particular instance that permeates most of the results exhibited in the rest of the paper happens when the linear space $\mathcal{V}$ is the dual of the Lie algebra $\mathfrak{h}^*$ of a Lie group $H$ and the orthogonal action of $H$ in $\mathfrak{h}^*$ (with respect to a given invariant metric) is the coadjoint action of $H$ on $\mathfrak{h}^*$.
We will denote by $\mathcal{O}_{x_0} = H\cdot x_0$ the coadjoint orbit of $H$ passing through $x_0 \in \mathfrak{h}^*$.  Let us recall that $\mathcal{O}_{x_0}$ carries a canonical symplectic structure.
Suppose now that $K \subset H$ is a compact subgroup of $H$, then the restriction to the coadjoint orbit $\mathcal{O}_{x_0} = H\cdot x_0$  of the canonical projection $\pi \colon \mathfrak{h}^* \to \mathfrak{k}^*$ is the momentum map of the action of $K$ in the symplectic manifold $\mathcal{O}_{x_0}$.

First, we can make a few simple remarks concerning the convex hull of the coadjoint orbit $H\cdot x_0 = \mathcal{O}_{x_0}$ and the range $\pi (\mathcal{O}_{x_0})$ of the momentum map.

\begin{lem}\label{conv_mom}  Under the assumptions above:

\begin{enumerate}

\item  $\pi (\mathrm{Conv}(\mathcal{O}_{x_0})) = \mathrm{Conv}(\pi (\mathcal{O}_{x_0}))$.

\item If $\mathrm{Conv} (H\cdot x_0)$ is a convex body, so is $\pi (\Conv(\mathcal{O}_{x_0}))$.

\end{enumerate}

\end{lem}
\textbf{Proof}
\begin{enumerate}
\item Because $\pi$ is a linear map, then $\pi
(x) \in \Conv(\pi(\mathcal{O}_{x_0}))$ for all $x \in
\Conv(\mathcal{O}_{x_0})$, hence
 $\pi (\mathrm{Conv}(\mathcal{O}_{x_0})) \subset \mathrm{Conv}(\pi
 (\mathcal{O}_{x_0}))$ and $\pi (\mathrm{Conv}(\mathcal{O}_{x_0}))$ is a convex
 set. Hence  $\pi (\mathrm{Conv}(\mathcal{O}_{x_0})) = \mathrm{Conv}(\pi
 (\mathcal{O}_{x_0}))$.
\item Suppose that $\pi (\Conv(\mathcal{O}_{x_0}))$ is not a convex body,
    hence because of Thm. \ref{th:main} there exist a fixed point $x$ and a
    proper subspace $W$ of $\mathfrak{k}^*$ such that $\pi
    (\Conv(\mathcal{O}_{x_0}))\subset x + W$.  Hence if we consider $y \in
    \Conv(\mathcal{O}_{x_0})$ such that $\pi (y) = x$ and $\widetilde{W} =
    \pi^{-1}(W^\perp)$ we have that $\Conv(\mathcal{O}_{x_0}) \subset y +
    \widetilde{W}$, and $\widetilde{W}$ is a proper subspace of
    $\mathfrak{h}^*$, hence because of Theorem~\ref{th:main},  $\Conv
    (H\cdot x_0)$ cannot be a convex body. \hfill $\square$
\end{enumerate}

In particular we may choose $K\subset H$ to be a maximal tori $T$, then $\pi\colon \mathfrak{h}^* \to \mathfrak{t}^* \cong \mathbb{R}^n$ with $n$ the rank of the group and  let $x_i \in \mathcal{O}_{x_0}$ be the fixed points of $T$.  It was observed by Kostant \cite{Ko73} that in such situation $\pi (\mathcal{O}_{x_0})$ is actually a convex polytope hence $\pi (\Conv(\mathcal{O}_{x_0})) = \Conv(\pi (\mathcal{O}_{x_0})) = \pi (\mathcal{O}_{x_0})$.  Now we can use Thm. \ref{th:main} to prove:

\begin{theo}\label{conv_coad}  Let $\mathcal{O}_{x_0} = K\cdot x_0 \subset \mathfrak{k}^*$ be a coadjoint orbit of the compact Lie group $K$ and $\pi \colon \mathcal{O}_{x_0} \to \mathfrak{t}^*$ the momentum map corresponding to the action of a maximal abelian subgroup $T$ of $K$, then if the convex hull $\Conv (K\cdot x_0 )$ is a convex body then $\pi(x_i)$ are not contained in a proper subspace of $\mathfrak{t}^*$ where $x_i$ are the fixed points of the action of $T$ in $\mathcal{O}_{x_0}$.
\end{theo}
\textbf{Proof}  As it was indicated before, because of Kostant and Atiyah's
convexity theorem \cite{At82}, the image of the momentum map $\pi \colon
\mathcal{O}_{x_0} \to \mathbb{R}^n$ is a convex polytope whose vertices are the
projections of the fixed points $x_i$ of the action of $T$ on
$\mathcal{O}_{x_0}$.  Then because of Lemma \ref{conv_mom} we have that the
convex hull $\Conv(K\cdot x_0)$ of the coadjoint orbit $\mathcal{O}_{x_0}$ is
just $\pi (\mathcal{O}_{x_0})$ and it is a convex body if $\Conv(K\cdot x_0)$
is. \hfill $\square$

By using Atiyah's convexity theorem \cite{At82} as indicated in the proof of
the previous theorem, or rather the extension of such theorem as proved by
Guillemin-Sternberg \cite{Gu82} and Kirwan \cite{Ki84} we can extend the result
in Theorem~\ref{conv_coad} as follows.   Let $M$ be a compact symplectic
manifold and $H$ a compact Lie group acting on it.  Let $J\colon M \to
\mathfrak{h}^*$ be the corresponding momentum map and $J(M) \mathfrak{h}^*$ its
range. Clearly $J(M)$ is a collection of coadjoint orbits of $H$.   Consider
the convex hull $\Conv{J(M)}$ of the range of the momentum map.  We can
characterize if it will be a convex body by using again a maximal abelian
subgroup $T \subset H$. Consider now $\mathfrak{t}^*$ embedded in
$\mathfrak{h}^*$ by using an invariant metric, then consider the intersection
of $J(M)$ with the positive Weyl chamber $\mathfrak{t}^*_+$.  According to
Guillemin-Sternberg-Kirwan's theorem,  $J(M)\cap \mathfrak{t}^*_+$ is a convex
polytope whose vertices are the fixed points of the action of $T$ \cite{Gu82}.
Hence we get:

\begin{cor}   If the convex hull $\Conv(J(M))$ of the family of coadjoint orbits $J(M)$ is a convex body then the fixed points of the action of $T$ are linearly independent.
\end{cor}

\section{Applications to Examples}\label{sec:applications}

\subsection{Convex body of density states}
To show how Corollary \ref{cmain} can be applied to seeing that mixed states form a convex body in
$\cA=u^*_1(\cH)$, consider first the orthogonal action of the unitary group $K=U(\cH)$ on the Euclidean space $\cV=u^*(\cH)$ of Hermitian operators on on a $d$-dimensional Hilbert space $\cH$, $d>1$, by

\be\label{rep}U.A=UAU^\dag\,.
\ee

\begin{prop}\label{p1}
The representation (\ref{rep}) of $U(\cH)$ in $u^*(\cH)$ has two irreducible
components: the space $\langle \mathbb{I}_\cH\rangle$, spanned by the
trace-normalized identity map
\[
\mathbb{I}_\cH=\frac{1}{d}I_\cH\,,
\]
and the subspace $su^*(\cH)$ consisting of all Hermitian operators with trace
0,
\be
\label{decomposition}
u^*(\cH) = \langle \mathbb{I}_\cH \rangle \oplus
su^*(\cH)\,.
\ee
\end{prop}
\textbf{Proof} The corresponding representation of the Lie algebra $su(\cH)$ in
$su^*(\cH)$ by $u.A=uA-Au=[u,A]$ is irreducible, as every invariant subspace
corresponds, {\it via} the multiplication by $i$, to a Lie ideal in the Lie
algebra $su(\cH)$ which is known to be simple. \hfill $\square$

If now $\ket{\psi}\in\cH$ is a nonzero vector, then the 1-dimensional
projector $P_{\psi}$ splits, according to (\ref{decomposition}), as

\[
P_{\psi}=\bI_\cH+\left(P_\psi-\bI_\cH\right)\,.
\]
Since the projection $\pi(P_\psi)$ onto $V(\cA)=su^*(\cH)$ is
$P_\psi-\bI_\cH\ne 0$ and $su^*(\cH)$ is irreducible, the set
$\cD(\cH)=\Conv(U(\cH).P_\psi)$ is a convex body in $u^*_1(\cH)$.Of course, the above constatation is well known and it is taken here to show
how Corollary \ref{cmain} works. Actually, more geometrical information is
known in this case. For instance, the radius of the largest ball $\cB$
contained in $\cD(\cH)$ and centred at $\bI_\cH$ is known (see
\cite{harriman78} or \cite[Corollary 3]{grabowski06}) to be
\be\label{radius}
r=\frac{1}{\sqrt{d(d-1)}}\,. \ee

This ball touches the boundary of $\cD(\cH)$
at points of the $U(\cH)$-orbit consisting of Hermitian operators with the
spectrum (diagonal form)

\[
\left(0,\frac{1}{d-1},\cdots,\frac{1}{d-1}\right)\,.
\]

\subsection{Convex body of separable states}
Let Hilbert spaces $\cH_1,\cH_2$ have dimensions $d_1,d_2>1$. A simple tensor
\[
\ket{\psi}=\ket{\phi^1}\ot\ket{\phi^2}\in\cH=\cH_1\ot\cH_2
\]
corresponds to a pure separable state $P_\psi=P_{\phi^1}\ot P_{\phi^2}$ whose
$K$-orbit under the obvious action of $K=U(\cH_1)\times U(\cH_2)\subset U(\cH)$
consists of all pure separable states, $K_.P_\psi=\cS^1(\cH)$. Its convex hull
is, by definition, the set $\cS(\cH)$ of all (mixed) separable states,
contained in the affine subspace $\cA=u^*_1(\cH_1\ot\cH_2)$ of
$\cV=u^*(\cH_1\ot\cH_2)=u^*(\cH_1)\ot_\R u^*(\cH_2)$. According to
Proposition~\ref{p1}, the decomposition of $\cV$ into irreducible parts is
\be\label{irrparts}
\fl\cV=\left(\la\bI_1\ra\ot\la\bI_2\ra\right)\oplus\left(\la\bI_1\ra\ot
su^*(\cH_2)\right)\oplus \left(su^*(\cH_1)\ot\la\bI_2\ra\right) \oplus
\left(su^*(\cH_1)\ot su^*(\cH_2)\right)\,,\ee where $\bI_j$ denotes
$\bI_{\cH_j}$, $j=1,2$. Here,

\be\label{decsu} \fl\left[\la\bI_1\ra\ot su^*(\cH_2)\right]\oplus
\left[su^*(\cH_1)\ot\la\bI_2\ra\right] \oplus \left[su^*(\cH_1)\ot
su^*(\cH_2)\right]=su^*(\cH_1\ot\cH_2)\,. \ee

The projection $\pi(P_\psi)=P_\psi-\bI_1\ot\bI_2$ of $P_\psi$ on
$su^*(\cH_1\ot\cH_2)$ decomposes as

\be\label{decomposition1}\bI_1\ot
(P_{\phi^2}-\bI_2)+(P_{\phi^1}-\bI_1)\ot\bI_2+(P_{\phi^1}-\bI_1)\ot(P_{\phi^2}-\bI_2)\,,
\ee
so all components in irreducible parts are non-trivial if $d_1,d_2>1$.
Hence, according to Corollary \ref{cmain},
$\Conv(K.P_\psi)=\cS(\cH_1\ot\cH_2)$ is a convex body in $u^*_1(\cH)$. Here
also more is known about the radius of the largest inscribed ball
\cite{gurvits02}.\subsection{Orbits of maximally entangled pure states}
For the composite system as above, assume that $d_1\ge d_2>1$ and take a unit
vector $\ket{\psi}\in\cH$. Decompose the projection

\[
\pi(P_\psi)=P_\psi-\bI_1\ot\bI_2\in su^*(\cH_1\ot\cH_2)
\]
into \be\label{psidec}\pi(P_\psi)=\bI_1\ot
P^{10}_\psi+P^{01}_\psi\ot\bI_2+P^{00}_\psi\,, \ee according to the
decomposition (\ref{decsu}) into irreducible parts. Then,

\[
\tr_1P_\psi-\bI_2=P^{10}_\psi\,,
\]
as $\tr_1(P^{01}_\psi\ot\bI_2+P^{00}_\psi)$ is clearly 0. If $P_\psi$ is
maximally entangled, then $\tr_1P_\psi-\bI_2$, thus $P^{10}_\psi$, is 0, so
$\pi(P_\psi)$ belongs to a proper $K$-invariant subspace and the convexed
orbit $\Conv(K.P_\psi)$ of $K=U(\cH_1)\ot U(\cH_2)$-action has empty interior
in $\cA=su^*_1(\cH_1\ot\cH_2)$.  Conversely, if the convexed orbit $\Conv(K.P_\psi)$ of $K=U(\cH_1)\ot
U(\cH_2)$-action has empty interior in $\cA=su^*_1(\cH_1\ot\cH_2)$, then
$\pi(P_\psi)$ belongs to a proper $K$-invariant subspace, so at least one of
$P^{10}_\psi, P^{01}_\psi,P^{00}_\psi$ is 0. Observe first that
$P^{00}_\psi\ne 0$ if only $d_2>1$. Indeed, in this case we can find
orthogonal $e_1,e_2\in\cH_1$ and $f_1,f_2\in\cH_2$ such that $\bk{e_1\ot
f_1}{\psi}_\cH\ne 0$, $\bk{e_2\ot f_2}{\psi}_\cH\ne 0$. But then

\[
A=\kb{e_1\ot f_1}{e_2\ot f_2}=\kb{e_1}{e_2}\ot\kb{f_1}{f_2}
\]
belongs to $su^*(\cH_1)\ot su^*(\cH_2)$, so

\[
\bk{P^{00}_\psi}{A}_{u^*}=\bk{P_\psi}{A}_{u^*}=\bk{\psi}{e_1\ot
f_1}_{\cH}\bk{e_1\ot f_1}{\psi}_\cH\ne 0\,.
\]
If $P^{01}_\psi=0$, then $\tr_2P_\psi=\bI_1$ and, according to Proposition
\ref{p0}, $P_\psi$ is maximally entangled. Finally, $P^{10}_\psi=0$ gives
$\tr_1P_\psi=\bI_2$ and, again, $P_\psi$ is maximally entangled. On the other
hand, as the $K$-action on $\cV_0=su^*(\cH_1)\ot su^*(\cH_2)$ is irreducible,
the orbit of a maximally entangled state is a convex body in $\bI_\cH+\cV_0$.
This proves the following characterization of maximally entangled states.

\begin{theo}\label{convmax} A pure state $P_\psi$ on $\cH_1\ot\cH_2$ is
maximally entangled if and only if the convexed orbit $\Conv(K.P_\psi)$ of the
canonical action of the group $K=U(\cH_1)\times U(\cH_2)$ in the space of
Hermitian operators on $\cH_1\ot\cH_2$ has empty interior in
$u^*_1(\cH_1\ot\cH_2)$ (so its volume in the convex body of density
states on $\cH_1\ot\cH_2$ is zero). In the latter case, however, $\Conv(K.P_\psi)$ is
a convex body in the affine space $\bI_\cH+su^*(\cH_1)\ot su^*(\cH_2)$.
\end{theo}

\subsection{The convex body of mixed-unitary channels}\label{muc}

As we already mentioned in section \ref{uc}, the set $\cC_{MUC}$ of
mixed-unitary channels is the convex hull of the orbit $\cO_{MUC}$ of the
channel $I_{gl}$ under the $K=U(\cH)\ti U(\cH)$-action (\ref{uaction}). We will
show that this picture is related, {\it via} the Jamio{\l}kowski  isomorphism
(\ref{jam}) to that in the previous section.It is well known that Hermiticity
preserving operators correspond, {\it via} the Jamio{\l}kowski  isomorphism, to
Hermitian operators on $\cH\ot\cH$. A convenient definition of $\wJ$ is given,
in the tensorial notation \cite{grabowski07}, by \be\label{jamdef} \left<
x_i\ot \bar x_j|A(x_k\ot\bar x_l) \right>=\left< x_i\ot x_l|\wJ(A)(x_j\ot x_k)
\right>\,, \ee Here, $x_i,x_j,x_k,x_l$ are arbitrary vectors in $\cH$ and
$x_i\ot\bar x_j$ is the tensorial notation for the Dirac's $\kb{x_i}{x_j}$. A
direct description in terms of a mixed tensorial-Dirac notation is the
following:

\be\label{jamdef1}
\wJ\left(\kb{x_i\ot \bar x_j}{x_k\ot\bar x_l}\right)=\kb{x_i\ot  x_l}{x_k\ot
x_j}\,. \ee

Here, $A=\kb{x_i\ot \bar x_j}{x_k\ot\bar x_l}$ represents
\be\label{A}A(\zr)= (x_i\ot \bar x_j)\circ\zr\circ(x_k\ot\bar
x_l)^\dag=(x_i\ot \bar x_j)\circ\zr\circ(x_l\ot\bar x_k)\,. \ee

From (\ref{jamdef}) one sees immediately that $A$ preserves positivity if and only if $\wJ(A)$ is positively defined:
\[
\left< x_i\ot \bar x_i|A(x_k\ot\bar x_k) \right>\ge 0 \Leftrightarrow\left< x_i\ot x_k|\wJ(A)(x_i\ot x_k) \right>\ge 0\,.
\]

The additional  doubly stochasticity conditions (\ref{stoch}) and (\ref{ds})
for (\ref{kraus}) correspond to the following conditions for partial traces:

\be\label{bsc} \tr_1\wJ(A)=I_\cH\,,\quad \tr_2\wJ(A)=I_\cH\,.
\ee Indeed, if
$(e_i)$ is an orthonormal basis in $\cH$, then

\[
\fl T_1\left(\kb{e_i\ot \bar e_j}{e_k\ot\bar e_l}\right)=(e_k\ot\bar
e_l)^\dag\circ(e_i\ot \bar e_j)=(e_l\ot\bar e_k)\circ(e_i\ot \bar
e_j)=\zd_k^i\cdot(e_l\ot \bar e_j),
\]
which coincides with
\[
\tr_1\left(\wJ\left(\kb{e_i\ot \bar e_j}{e_k\ot\bar
e_l}\right)\right)=\tr_1\left(\kb{e_i\ot  e_l}{e_k\ot
e_j}\right)=\zd_k^i\cdot(\kb{e_l}{e_j})\,.
\]
Similarly, \be\label{Tt} T_2(A)=\tr_2\wJ(A)\,. \ee This means that $\wJ$
establishes an isomorphism between the convex set of doubly stochastic
operators and the convex set of those non-negatively defined operators on
$\cH\ot\cH$ whose both partial traces equal $I_\cH$.Another important
observation is that $\wJ$ intertwines the $K=U(\cH)\ti U(\cH)$-action
(\ref{uaction}) on $HP(gl(\cH))$ with the standard $K$-action (\ref{newaction})
on $u^*(\cH\ot\cH)=u^*(\cH)\ot u^*(\cH)$. Indeed, for $A$ as in (\ref{A}), it
is easy to see that
\beas \fl((U_1,U_2).A)(\zr)&=&U_1A(U_2\zr U_2^\dag)
U_1^\dag=
U_1\circ(x_i\ot \bar x_j)\circ U_2\circ\zr\circ U_2^\dag\circ(x_l\ot\bar x_k)\circ U_1^\dag\\
&=&(U_1x_i\ot\overline{U_2x_j})\circ\zr\circ(U_2x_l\ot\overline{U_1x_k})\,,
\eeas
so that
\bea\label{intertw} \fl &\wJ\left((U_1,U_2).(\kb{x_i\ot \bar
x_j}{x_k\ot\bar x_l})\right)=\wJ(
\kb{U_1x_i\ot\overline{U_2x_j}}{U_1x_k\ot\overline{U_2x_l}}) \nonumber \\
\fl &=\kb{U_1x_i\ot U_2x_l}{U_1x_k\ot{U_2x_j}})=(U_1\circ\kb{x_i}{x_k}\circ
U_1^\dag)\ot(U_2\circ\kb{x_l}{x_j}\circ U_2^\dag)\,.
\eea

All this implies that our convex set $\cC_{MUC}$ is Jamio{\l}kowski  equivalent
to the convex hull $\Conv(\cO)$ of the orbit $\cO=K.\wJ(I_{gl})$. But,

\[
\wJ(I_{gl})=\wJ\left(\sum_{i,j}\kb{e_i\ot \bar e_i}{e_j\ot\bar e_j}\right)=
\kb{\sum_{i}e_i\ot e_i}{\sum_{j}e_j\ot e_j}\,,
\]

where $(e_i)$ is an orthonormal basis in $\cH$. The latter, however, is
proportional to a maximally entangled pure state $P_{\psi}$ associated with
the normalized vector

\[
\ket{\psi}=\frac{1}{\sqrt{\dim(\cH)}}\sum_ie_i\ot e_i\,.
\]

More precisely,

\be\label{max} \wJ(I_{gl})=\dim(\cH)\cdot P_\psi\,. \ee

Now, we are in the situation of the previous section; the only difference is that
all is rescaled by $\dim(\cH)$. In view of Theorem \ref{convmax}, the convex
hull of the $K$-orbit of $\wJ(I_{gl})$ is then a convex body in the affine
space

\[
\cA=\dim(\cH)\cdot\bI_{\cH\ot\cH}+su^*(\cH)\ot su^*(\cH)\,.
\]

In consequence, $\cC_{MUC}$ is a convex body inside the set of doubly
stochastic channels. The convex body $\cC_{MUC}$ is clearly centred at

$$\zW=\dim(\cH)\cdot\wJ^{-1}(\bI_{\cH\ot\cH})=\frac{1}{\dim(\cH)}\cdot\wJ^{-1}(I_{\cH\ot\cH})\,.$$

But, according to (\ref{jamdef}),

\[
\left< e_i\ot \bar e_j|\zW(e_k\ot\bar e_l)
\right>=\frac{1}{\dim(\cH)}\cdot\zd_i^j\zd_k^l\,,
\]
which immediately implies that

\be\label{cdc}
\zW(X)=\frac{\tr(X)}{\dim(\cH)}I_\cH=\tr(X)\bI_\cH\,.
\ee

The mixed-unitary channel $\zW$ is called the {\it completely depolarizing
channel}.One can find $\zW$ easily also without the use of Jamio{\l}kowski
isomorphism. It is clear that

\be\label{centred} \zW(X)=\int_{U(\cH)}UX U^\dag d\zm(U)\,, \ee where $\zm$ is
the probabilistic Haar measure on $U(\cH)$. Since $\zW$ is stabilized by
$U(\cH)$,
\[
U\zW(X)U^\dag=\zW(X)\,,
\]
for any Hermitian $X$ and any $U\in U(\cH)$. This implies that $\zW(X)$ is
proportional to $I_\cH$, i.e., \be\label{igl3} \zW(X)=\tr(X_0X)I_\cH \ee for a
certain $X_0\in u^*(\cH)$.On the other hand, any left-invariant Haar measure on
$U(\cH)$ is automatically right-invariant, so
\[
\zW(UXU^\dag)=\zW(X)
\]
and thus $\tr(X_0UXU^\dag)=\tr(X_0X)$ for all $X$ and all $U$. Hence, $X_0$
is proportional to the identity, $X_0=c\cdot I_\cH$ and
\[
\zW(X)=c\cdot\tr(X)I_\cH\,.
\]
Finally, as $\zW(I_\cH)=I_\cH$, we get $1=c\cdot\dim(\cH)$, thus (\ref{cdc}).
We can summarize as follows.

\begin{theo}
Any doubly stochastic channel in a neighborhood of the completely
depolarizing channel $\zW$ is mixed-unitary.
\end{theo}

This is clearly a slightly weaker version of a recent result of  Watrous
\cite{watrous08}.

\subsection{The largest balls of $k$-entangled states}
Consider again a bipartite Hilbert space $\cH=\cH_1\ot\cH_2$ of the total
dimension $d=d_1d_2$, where $d_1=\dim(\cH_1)\ge\dim(\cH_2)=d_2$, and consider
the convex sets $\cE_k(\cH_1\ot\cH_2)$ of $k$-entangled states,
$k=1,2,\dots,d_2$.It is known \cite{gurvits02} that the radius of the largest
ball contained in $\cS(\cH_1\ot\cH_2)$ and centered at $\bI_\cH=\bI_1\ot\bI_2$
is $r=\frac{1}{\sqrt{d(d-1)}}$, with $d=d_1d_2$. This is exactly the same ball
as the largest ball $\cB$  (\ref{radius}) contained in the (bigger) convex body
$\cD(\cH)$ of all mixed states (see \cite{harriman78, grabowski06}). In other
words, $\bI_\cH+A$ is separable for all $A$ with $\nm{A}_{u^*}\le a$ if and
only if $a\le\frac{1}{\sqrt{d(d-1)}}$.This observation, however, implies
immediately that the largest ball $\cB_k$, centered at $\bI$ and contained in
$\Conv(\cE_k(\cH_1\ot\cH_2))$, must be the same, since

\[
\cS^1(\cH_1\ot\cH_2)=\cE_1(\cH_1\ot\cH_2)\subset\cE_k(\cH_1\ot\cH_2)\subset\cD^1(\cH)\,.
\]

\begin{prop}
The largest ball $\cB_k$, centred at $\bI$ and contained in
$\Conv(\cE_k(\cH_1\ot\cH_2))$ has radius (\ref{radius}) and coincides with the
largest ball $\cB$ contained in the convex body $\cD(\cH)$ of all density
states, for all $k=1,2,\dots,d_2$. In particular, $\bI_\cH+A$ is
$k$-entangled for all $A$ with
\[
\nm{A}_{u^*}\le \frac{1}{\sqrt{d(d-1)}}\,.
\]
\end{prop}


\section{Convexed local orbits}

As the $k$-entangled states are convex hulls of families of orbits, in spite of
the above proposition, looking for single orbits of a particular pure bipartite
state is still an interesting problem.

Let $\ket{\psi}\in\cH=\cH_1\ot\cH_2$ be a nonzero vector, $k=\Sr(\psi)$ be its
Schmidt rank, and $P_\psi=\frac{\kb{\psi}{\psi}}{\bk{\psi}{\psi}}$ be the
corresponding pure state.With $C_\psi$ we will denote the {\it convexed local
orbit} of $\zr=P_\psi$, i.e. the convex hull of the orbit $O_\psi=K.P_\psi$ of
the pure state $P_\psi$ under the unitary action $\zr\mapsto U.\zr=U\zr U^\dag$
of the group $K=U(\cH_1)\ti U(\cH_2)$, where $U$ runs over all local unitary
operators $U\in U(\cH_1)\ti U(\cH_2)$ represented by the tensor products
$U_1\ot U_2$, $U_i\in U(\cH_i)$, $i=1,2$. According to the Schmidt
decomposition (\ref{Sd}) and the form of the partial trace
(\ref{partialtrace}), elements $\zr$ in the orbit $O_\psi$ are determined by
the spectrum $(\zl_1^2,\dots,\zl_k^2)$ of their partial trace $\tr_1\zr$.
Indeed, the spectrum determines $\zl_1,\dots,\zl_k>0$ and thus the Schmidt
decomposition (\ref{Sd}) which identifies the pure state up to a local unitary
transformation.

\begin{theo}\label{tt}
The convexed local orbit $C_\psi$ is a $K$-invariant subset of $u^*_1(\cH)$
centred at $\bI_\cH$ and contained in the convex body $\cD(\cH)$ of all density
states. Moreover, $C_\psi$ is itself a convex body unless $\ket{\psi}$ is
maximally entangled.
\end{theo}
\textbf{{Proof}} In view of Theorem \ref{convmax}, it is enough to show that
$\bI_\cH\in C_\psi$. Take the probabilistic Haar measure $\zm$ on
$K=U(\cH_1)\ti U(\cH_2)$ and consider
\[
\zr_0=\int_KUP_\psi U^\dag d\zm(U)\in u^*_1(\cH)\,.
\]
By construction, $\zr_0$ is a $K$-invariant element in $C_\psi$. It is easy
to see that $\zr_0=\bI_\cH$. Indeed, using decomposition (\ref{psidec}), we
get
\[
\zr_0-\bI_1\ot\bI_2=\int_KU.\left(\bI_1\ot
P^{10}_\psi+P^{01}_\psi\ot\bI_2+P^{00}_\psi\right)d\zm(U)=0\,,
\]
since the latter integral reduces to
\beas &\bI_1\ot\int_{U(\cH_2)}
U_2P^{10}_\psi U_2^\dag\, d\zm_2(U_2)+\int_{U(\cH_1)} U_1P^{01}_\psi U_1
^\dag\, d\zm_1(U_1)\ot \bI_2\\&+ \sum_j\left(\int_{U(\cH_1)} U_1P^{1}_j U_1
^\dag\, d\zm_1(U_1)\ot\int_{U(\cH_2)} U_2P^{2}_j U_2^\dag\,
d\zm_2(U_2)\right) \eeas
and the only $U(\cH_i)$-invariant element in
$u^*_1(\cH_i)$ is 0. Here, $\zm_i$ is the probabilistic Haar measure on
$U(\cH_i)$, $i=1,2$, and

\[
P^{00}_\psi=\sum_j\left(P^1_j\ot P^2_j\right)\,.
\]
\hfill $\square$

\section{Maximum volume ellipsoids}\label{sec:ellipsoids}
Let us recall that among all ellipsoids contained in a convex body $C$ there
is a unique ellipsoid $E_{max}(C)$ of the maximum volume, which we call the
{\it maximum volume ellipsoid} of $C$ and which is also called the {\it John
ellipsoid} of $C$ \cite{john48}. Actually, $E_{max}(C)$ does not depend on
the choice of an Euclidean metric in $C$, so it is determined completely by
the affine (and convex) structure. On the other hand, it is clear that
$E_{max}(C)$ may be larger than the largest ball $\cB(C)$ contained in $C$,
since the latter clearly depends strongly on the metric. However, in many
important cases of convex bodies in Euclidean spaces the maximal ellipsoids
are largest balls. For instance, this is the case of the convex body
$\cD(\cH)$ of all density states that easily follows from the following
observation.

\begin{prop}\label{p11} If a compact group $K$ acts irreducibly on an Euclidean
space $\cV$ by orthogonal transformations, then the maximum volume ellipsoid
contained in the convex hull $C=\Conv(K\cdot x_0)$ of any $K$-orbit is a ball,
$E_{max}(C)=\cB(C)$.
\end{prop}
\textbf{Proof} We may assume that $x_0\ne 0$, so that $C$ is a convex body in
$\cV$ centred at 0. We will show that the largest ball $\cB$ centred at 0 and
contained in $C$ coincides with $E_{max}$. Indeed, $\cB\subset E_{max}$ and it
suffices to show that all principal axes of $E_{max}$ are equal. Suppose the
contrary and let $v\in\cV$ be the direction of the largest axis. Let $\cV_0$ be
the orthogonal completion of $v$. As the boundary of $\cB$ intersects the
boundary of $E_{max}$ in $\cV_0$, the only points at which $\cB$ touches the
boundary of $C$ must lie in $\cV_0$. But these point form a $K$-invariant
subset, thus span a proper $K$-invariant subspace in $\cV$; a contradiction
with the irreducibility. \hfill $\square$

As $C=\cD(\cH)-\bI_\cH$ is the convex hull of an orbit of $U(\cH)$-action on
$su^*(\cH)$, $E_{max}(\cD(\cH))=\cB(\cD(\cH))$. This is, however, no longer
true for convexed local orbits $C_\psi$ of pure states in
$\cH=\cH_1\ot\cH_2$.Let us consider the simple case of a two-qubit system:
$\dim(\cH_1)=\dim(\cH_2)=2$. Suppose that a normalized vector
$\ket{\phi}\in\cH$ has a Schur-like decomposition
\[
\ket{\phi(\zl)}=\zl\cdot e_1\ot f_1+\sqrt{1-\zl^2}\cdot e_2\ot f_2\,,
\]
with $0\le\zl^2\le 1$. Here, $(e_1,e_2)$ and $(f_1,f_2)$ are orthonormal bases
in $\cH_1$ and $\cH_2$, respectively.If $\zl^2$ varies from 1 to $1/{2}$ (or
from 0 to $1/{2}$), then $P_{\phi(\zl)}$ varies from a separable to the
maximally entangled pure state $P_{{\psi}}=P_{\phi(\pm 1/\sqrt{2})}$ associated
with
\[
\ket{\psi}=\frac{1}{\sqrt{2}}\left(\pm e_1\otimes f_1+e_2\otimes f_2\right).
\]
Let $R(\zl)$ be the radius of the largest ball $\cB(\zl)$ centred at $\bI_\cH$
and contained in $$C(\zl)=\Conv\left(K.P_{\phi(\zl)}\right)\,,\quad
K=U(\cH_1)\ti U(\cH_2)\,.$$ According to Theorem \ref{tt}, $C(\zl)$ is a convex
body in $u^*_1(\cH)$ if and only if $\zl\ne\frac{1}{\sqrt{2}}$. If
$\zl=\frac{1}{\sqrt{2}}$, then $C(\zl)-\bI_\cH$ flattens to a convex body in
the irreducible subspace $su^*(\cH_1)\ot su^*(\cH_2)$.

In view of Proposition \ref{p11}, the largest ball $\cB(1/\sqrt{2})$ is the
maximal volume ellipsoid. We will show that this is not true in general, i.e.,
$E_{max}(C(\zl))$ differs from $\cB(\zl)$ for $\zl^2$ close to $1/2$, $\zl^2\ne
1/2$.The partial traces of $P_{\phi(\zl)}$ are:

\[
\tr_1P_{\phi(\zl)}=\zl^2P_{f_1}+(1-\zl^2)P_{f_2}\,,\quad
\tr_2P_{\phi(\zl)}=\zl^2P_{e_1}+(1-\zl^2)P_{e_2}\,,
\]
so that, in the decomposition (\ref{psidec}),
\[
\fl P^{10}_{\phi(\zl)}=\left(\zl^2-\frac{1}{2}\right)P_{f_1}-\left(\zl^2-\frac{1}{2}\right)P_{f_2}\,,
\quad
P^{01}_{\phi(\zl)}=\left(\zl^2-\frac{1}{2}\right)P_{e_1}-\left(\zl^2-\frac{1}{2}\right)P_{e_2}\,.
\]

This implies that the orthogonal projection of $C{(\zl)}-\bI$, thus of
$\cB(\zl)-\bI$, onto the subspace $\la\bI_1\ra\ot su^*(\cH_2)$ lies in the
ball of the radius

\[
r(\zl)=\nm{\bI_1\ot\left(\tr_1P_{\phi(\zl)}-\bI_2\right)}=\zl^2-\frac{1}{2}\,.
\]

Hence, $R(\zl)\le r(\zl)$. Since $\cB(\zl)\subset\cB(\cD(\cH))$ and the
latter has the radius $\frac{1}{\sqrt{12}}$ (cf. (\ref{radius})), we get the
following.

\begin{prop} The radius of $\cB(\zl)$ can be estimated by
\[
R(\zl)\le min\left\{\zl^2-\frac{1}{2},\frac{1}{\sqrt{12}}\right\}\,.
\]
In particular, $R(\zl)\to 0$ as $\zl^2\to\frac{1}{{2}}$.
\end{prop}

Let us note that, given $\zl$, both states $P_{\phi(\pm\zl)}$ belong to the same $K$-orbit, so $C(\zl)=C(-\zl)$ and

\be\label{xx}
\zr_0=\frac{1}{2}\left(P_{e_1}\ot P_{f_1}+P_{e_2}\ot P_{f_2}\right)=\frac{1}{2}\left(P_{\phi(\zl)}+ P_{\phi(-\zl)}\right)
\ee
belongs to $C(\zl)$ for all $-1\le\zl\le 1$. In particular, $\zr_0$ lies in the convexed orbit of maximally entangled states, so that $\zr_0-\bI\in su^*(\cH_1)\ot su^*(\cH_2)$ and we get the following.

\begin{prop}
The convexed $K$-orbit $C_0=\Conv(K.\zr_0)$ is a convex body in the affine space $\cA_0=\bI+su^*(\cH_1)\ot su^*(\cH_2)$, contained in $C(\zl)$ for any $-1\le\zl\le 1$.
\end{prop}

If now $\cB_0=\cB(C_0)$ is the largest ball in $C_0$ and $r_0$ is the radius
of $\cB_0$, then $\cB(\zl),\cB_0\subset C(\zl)$. Hence,
$\cB(\zl)/2+\cB_0/2\subset C(\zl)$. In particular,

\[
\fl\bI+\zr^{10}+\zr^{01}+\zr^{00}\in \bI+\left(\la\bI_1\ra\ot
su^*(\cH_2)\right)\oplus \left(su^*(\cH_1)\ot\la\bI_2\ra\right) \oplus
\left(su^*(\cH_1)\ot su^*(\cH_2)\right)
\]
belongs to $C(\zl)$ if only $\nm{\zr^{10}+\zr^{01}}\le \frac{R(\zl)}{2}$ and
$\nm{\zr^{00}}\le \frac{r_0}{2}$. This implies the following.

\begin{theo}
The ellipsoid
\[
E(\zl)=\left\{
\bI+\zr^{10}+\zr^{01}+\zr^{00}:\frac{\nm{\zr^{10}}^2}{R(\zl)^2}+\frac{\nm{\zr^{01}}^2}{R(\zl)^2}+
\frac{\nm{\zr^{00}}^2}{r_0^2}\le\frac{1}{4}\right\}
\]
is contained in $C(\zl)$. The fraction of volumes,
\[
\frac{vol(B(\zl))}{vol(E(\zl))}=\left(\frac{R(\zl)}{r_0}\right)^9
\le\left(\frac{\zl^2-\frac{1}{2}}{r_0}\right)^9\,,
\]
tends to 0 as $\zl^2\to 1/2$. In particular, $\cB(\zl)\ne E_{max}(\zl)$ for
$\zl^2$ close to $\frac{1}{2}$, $\zl^2\ne \frac{1}{2}$.
\end{theo}

\section{Summary}
Numerous problems of quantum information theory involve convex combinations of
linear operators taken form a prescribed set. The most prominent example is
that of mixed separable density states which, form definition, are convex
combinations of pure separable states, i.e.\ simple tensor products of
projections on one-dimensional subspaces of the underlying Hilbert space. In
processes of transformation and transmission of quantum information, one is
often confronted with possibilities of applying several quantum channels with
some prescribed probabilities, what again leads to convex combinations of
operators representing channels.  Usually, convex sets obtained in this manner
are of practical importance only if they constitute a significant part of the
whole set of states or channels, i.e. when they form a \textit{convex body} in
these sets, or in other words, contain an open subset of the set of all states
or channels. In the paper, we gave a unifying way of deciding whether this is
the case when the set in question is a convexed orbit of some symmetry group
through some distinguished state(s) or channel(s). This is a fairly general
situation, since usually we have at our disposal the local symmetry group
consisting of invertible quantum operations applied individually to components
of a composite quantum system. The general problem (see Problem~\ref{pr1}),
whether the convex hull of an orbit is a convex body is answered by
Theorem~\ref{th:main}, which is then applied to various cases involving state
and channels. In particular, we gave a unique characteristic of maximally
entangled states (see Theorem~\ref{convmax}) in terms of the convexed orbits of
the local group through them.   A state is maximally entangled if the convex
hull of the orbit has an empty interior in the space of all density states of a
composite system.   The characterization of orbits whose convex hulls are
convex bodies provided by Theorem~\ref{th:main} combined with the
Atiyah-Guillemin-Sternberg-Kirwan's theorem on the convexity properties of the
momentum map is applied to the study of the convex hull of coadjoint orbits,
showing that they are convex bodies if an independence properties of the fixed
points of the action of a Cartan subgroup is satisfied.

Convex bodies can be partially characterized by the largest balls around some
distinguished ``center" contained in the body, Such a characterization is
useful when analyzing how strong we may perturb the distinguished (e.g.\
maximally separable) state without loosing a desired property (e.g.\
separability). Such a characterization depends on the metric used. There
exists another way of portraying a convex body in an approximate way in terms
of the \textit{maximum volume ellipsoid} contained in it, which is actually
independent on the choice of metric, bearing thus purely affine character.
The largest ball and the maximal ellipsoid can, however, coincide in some
cases (for a particular choice of a ``natural" metric) and differ in other
cases. We showed that the former situation occurs for the convex body of
states embedded in the space of trace-one operators, whereas the latter takes
place for convexed local orbits through pure states of bipartite systems. In
both cases the natural metric is the Hilbert-Schmidt one.




\ack

We thank Karol \.Zyczkowski for suggesting some of the problems considered in
the paper. Research of J.~Grabowski and M.~Ku\'s was financed by the Polish
Ministry of Science and Higher Education under the grant No. N N202 090239.
G.~Marmo would like to acknowledge the support provided by the
Santander/UCIIIM chair of Excellence programme 2011-2012.  A. Ibort
would like to acknowledge the partial support provided by the Fundaci\'on Caja Madrid
and MICIN project  MTM2010-21186-C02-02.



\end{document}